\documentclass[ ]{revtex4-2}
\usepackage{graphics,epsfig}
\usepackage{epstopdf}
\usepackage{graphicx}
\usepackage{dcolumn}
\usepackage{amsmath}
\usepackage{epstopdf}
\usepackage{amsmath, amssymb, amsthm, graphicx, hyperref, geometry}
\begin{document}
\title{Holography and the Swampland: Constraints on Quantum Gravity from Holographic Principles}
\author{Sudhaker Upadhyay\footnote{Visiting Associate, Inter-University Centre for Astronomy and Astrophysics (IUCAA) Pune-411007, Maharashtra, India}}
\email{sudhakerupadhyay@gmail.com}
\affiliation{Department of Physics, K.L.S. College, Nawada, Magadh University, Bodh Gaya, Bihar 805110, India}
\author{Alexander A. Reshetnyak}
\email{reshet@tspu.ru}

 \affiliation{Center for Theoretical Physics, Tomsk State Pedagogical University, Tomsk 634061, Russia, \\
  National Research Tomsk Polytechnic   University,
 Tomsk 634050, Russia}
 \author{Pavel Yu. Moshin}
\email{pavel.moshin@mail.ru}

 \affiliation{Núcleo Interdisciplinar de Ciências Exatas e da Natureza, NICEN, Universidade Federal de Pernambuco, Centro Acadêmico do Agreste, CAA, Brazil}

  \author{Ricardo A. Castro}
\email{rialcap@usp.br}

 \affiliation{Núcleo Interdisciplinar de Ciências Exatas e da Natureza, NICEN, Universidade Federal de Pernambuco, Centro Acadêmico do Agreste, CAA, Brazil}
 \affiliation{Department of Nuclear Physics, Institute of Physics, University of São Paulo, São Paulo CEP 05508-090, Brazil}
\begin{abstract}
The Swampland Program aims to delineate the space of consistent low-energy effective field theories (EFTs) that admit a UV completion in quantum gravity from those that do not. In parallel, holography, and particularly the AdS/CFT correspondence, offers a non-perturbative definition of quantum gravity in asymptotically anti-de Sitter (AdS) spacetimes. In this paper, we explore the Swampland Conjectures through the lens of holography, focusing on how holographic consistency conditions, such as the convexity of the conformal field theory (CFT) spectrum, the averaged null energy condition (ANEC), and the modular bootstrap, map onto Swampland constraints in the bulk. We argue that the holographic principle provides a geometric realization of Swampland bounds, particularly on scalar field potentials and the absence of long-lived de Sitter vacua. Finally, we discuss how the emergent bulk locality in AdS/CFT provides evidence that the Swampland conjectures may themselves be manifestations of deeper holographic consistency conditions.
\end{abstract}
\maketitle
\section{Introduction}

Over the past two decades, the quest to understand the nature of quantum gravity has led to two complementary conceptual frameworks: the Swampland program and the holographic principle. Both approaches aim to characterize what distinguishes consistent theories of quantum gravity from effective field theories (EFTs) that appear consistent at low energies but cannot arise from a UV-complete gravitational theory. While the Swampland program is motivated primarily by internal consistency conditions of string theory and the structure of moduli spaces, holography provides a concrete non-perturbative realization of quantum gravity via the AdS/CFT correspondence. These frameworks give us a profound glimpse into how quantum spacetime is structured and what rules might shape it.

The idea of the landscape and swampland was first articulated by Vafa \cite{Vafa2005}, who proposed that only a small subset of effective field theories—those compatible with a consistent embedding into string theory—constitute the so-called landscape, while all others reside in the swampland. This perspective was sharpened by Ooguri and Vafa \cite{OoguriVafa2007}, who argued that certain universal constraints—such as bounds on scalar field excursions, the absence of exact global symmetries, and the requirement that gravity be the weakest force—are likely necessary conditions for any EFT to have a consistent quantum-gravitational completion. These conjectures, collectively known as the Swampland Conjectures, have since evolved into a broad research program with deep implications for cosmology, particle physics, and the structure of spacetime \cite{Palti2019, Brennan2017, vanBeest2021,   1, 2, 3, 4, 5}. Swampland conjectures  have found direct consequences for formation of primordial
black holes   and dark matter  \cite{05}. Recently,  the compatibility of curvature-matter coupling gravity theory within the framework of swampland conjectures is studied \cite{06}. The recent progress on this subject can be found in Refs.    \cite{Chameleons2024,09,010,011,0012}.

Recent investigations continue to refine how quantum-gravity consistency constrains low-energy effective theories across cosmology and black-hole physics. The analysis of charged black holes with perfect-fluid dark matter in \cite{2411.04134} demonstrates that WGCConjecture (WGC)–driven particle emission can enforce the preservation of cosmic censorship near extremality. Complementarily, \cite{2404.15981} shows that scalar-field cosmologies with curved field spaces can satisfy both the Trans-Planckian Censorship and Swampland Distance Conjectures while remaining compatible with late-time acceleration. Earlier work such as \cite{2205.03648} sharpens the general structure of swampland constraints on effective field theories, clarifying the limitations imposed by ultraviolet completion. Moreover, \cite{2406.13784} highlights that these restrictions extend to broader gravitational settings, influencing black-hole backgrounds and their thermodynamic properties. Collectively, these studies underscore the growing precision of the swampland program as a framework for assessing which theoretical models can plausibly arise from a consistent theory of quantum gravity.

Among the most influential are the Distance Conjecture, which predicts the exponential lightening of an infinite tower of states as a scalar field traverses a trans-Planckian distance in moduli space \cite{OoguriVafa2007, Klaewer2017, Grimm2018}; the WGC, asserting that gravity must be the weakest force \cite{ArkaniHamed2006, Heidenreich2015, Montero2016, 6}; and the de Sitter Conjecture, which challenges the existence of stable or metastable de Sitter vacua in consistent quantum gravity \cite{Obied2018, Garg2019, Ooguri2018}. While these conjectures remain speculative, mounting evidence from string compactifications, black hole thermodynamics, and dualities suggests they capture essential features of the gravitational UV completion. 

On the other hand, the holographic principle provides an entirely different lens on the problem of quantum gravity. Originally proposed by ’t~Hooft and Susskind \cite{tHooft1993, Susskind1995}, holography posits that the number of degrees of freedom in a gravitational system scales with its boundary area rather than its volume. This principle found its most precise realization in the AdS/CFT correspondence \cite{Maldacena1998, Witten1998, Gubser1998}, which equates a $d$-dimensional  CFT  to a $(d+1)$-dimensional theory of gravity in asymptotically  AdS  spacetime. Holography has since become one of the central tools for understanding the dynamics of quantum gravity, offering explicit dictionary entries between bulk fields and boundary operators, between bulk geometry and entanglement structure, and between black hole thermodynamics and quantum information \cite{Hubeny2007, VanRaamsdonk2010, Faulkner2014, Harlow2018, RyuTakayanagi2006}.

From a holographic standpoint, the Swampland program acquires a new and potentially illuminating interpretation. In AdS/CFT, every consistent bulk theory must correspond to a unitary, local, and crossing-symmetric CFT with a positive-definite spectrum of operator dimensions \cite{Heemskerk2009, Fitzpatrick2011, Hartman2015}. The consistency of the CFT thus encodes the consistency of the bulk gravitational EFT. The space of consistent CFT data can therefore be viewed as the holographic landscape, while the remaining bulk theories—those that fail to admit a unitary boundary dual—constitute the holographic swampland \cite{Baumann2020, Andriolo2018}. In this framework, the Swampland Conjectures may be understood as emergent consequences of holographic consistency conditions such as unitarity, analyticity, causality, and the positivity of energy \cite{Hartman2009, Komargodski2016, Camanho2016}.

Moreover, holography provides a natural explanation for certain Swampland bounds. For instance, the WGC can be interpreted as a requirement that charged black holes in AdS be unstable to decay into lighter states, which corresponds in the boundary theory to the existence of charged operators saturating the BPS-like bound $\Delta \leq q L$ \cite{NakayamaOoguri2016}. Similarly, the Distance Conjecture manifests in the dual CFT as the accumulation of light operators when moving along a non-compact conformal manifold, signaling the breakdown of bulk locality \cite{Grimm2018, Lee2019, Gonzalo2019}. Even the absence of stable de Sitter vacua has been argued to follow from holographic reasoning: a Euclidean CFT dual to a stable de Sitter space would violate reflection positivity and modular invariance, suggesting such vacua are inconsistent with holography \cite{Harlow2011, Anninos2012, Maldacena2019}.

{In this connection,  the~extension of General Relativity on a base of local supersymmetry principle up to the supergravity models~\cite{Nieuwenhuizen} with improved quantum properties and a connection with (super)string field theory permits one to include massless fields of spins $s>2$ in higher spin gravity (see~\cite{Snowmass} and references therein) with respecting the string field theory properties, asymptotic freedom and some others. The~mentioned AdS/CFT correspondence gives valuable indications that higher spin excitations can be significant to elaborate the quantum gravity challenges~\cite{Giombi}. Some higher spin models for free and interacting (half-)integer massless and massive fields within Lagrangian dynamics on AdS spaces  \cite{0607248_AdS_amb, frame-like1, BKR, RYT2}   may additionally served  provide another  possible
insight into the origin of dark matter and dark energy, see for reviews~\cite{DM1,DM2,Odintsov1}.}

In this paper, we develop this correspondence systematically. We argue that the Swampland constraints can be reinterpreted as statements about the internal consistency of holographic theories, and that many of the conjectures acquire a natural explanation from the geometric and entropic structure of holographic dualities. In particular, we focus on how the emergence of bulk locality, the convexity of the CFT spectrum, and the positivity of relative entropy provide a unified framework linking holographic consistency to Swampland bounds. Ultimately, we propose that the Swampland program and holography are not separate endeavors but two complementary expressions of a single principle: the consistency of quantum gravity as seen from different vantage points.

Although our analysis is formulated within the AdS holographic framework, it is still meaningful to discuss the de Sitter Swampland Conjecture. Our focus on this conjecture stems from the fact that holographic consistency conditions, such as boundary positivity, modular stability, and constraints on bulk potentials, naturally generate bounds that parallel those appearing in the de Sitter case. In this sense, the conjecture is invoked not to assume a de Sitter background, but to highlight how holographic arguments restrict the existence of positive-energy vacua in any putative quantum-gravity completion. The AdS Swampland statements operate at a complementary level, constraining admissible AdS vacua, whereas the de Sitter conjecture serves here as a probe of how holographic principles limit the construction of stable or metastable vacua with positive potential energy.

Throughout this work we adopt the electric formulation of the WGC, together with the widely used tower-based refinement. In this version, the conjecture asserts the existence of charged states whose charge-to-mass ratio exceeds that of an extremal black hole, and moreover that such states appear as part of an infinite tower whose masses decrease as one approaches weak-coupling limits. We employ this formulation because it is the one most naturally aligned with holographic consistency requirements: in particular, with the convexity properties of charged operator dimensions in the boundary CFT and with the emergence of light charged spectra in weakly coupled regions of moduli space. No assumption about a specific microscopic realization is required; what matters for our purposes is that the conjecture be interpreted as a constraint on the spectrum of charged excitations rather than a statement tied to a single isolated superextremal particle.
\section{Swampland Conjectures and Quantum Gravity}

The Swampland program aims to delineate the boundary between effective field theories that can arise as low-energy limits of consistent quantum gravity and those that cannot. Unlike the traditional bottom-up approach to model building, which starts from low-energy consistency requirements such as locality, unitarity, and gauge invariance, the Swampland perspective introduces new constraints rooted in ultraviolet (UV) completion. These conjectures emerge from recurring patterns observed across a broad range of string compactifications, black hole physics, and general properties of quantum gravitational theories \cite{Vafa2005, OoguriVafa2007, Palti2019}. In this section, we review and extend the mathematical formulation of the central conjectures, highlighting their internal consistency and their connections through the geometry of field space and quantum gravity dynamics.

\subsection{The Distance Conjecture and Moduli Space Geometry}

The Distance Conjecture posits that when a scalar field $\phi$ in a quantum gravitational theory moves over a large distance in moduli space—typically of order $\Delta \phi \gtrsim M_P$—an infinite tower of states becomes exponentially light. Quantitatively, one writes
\begin{equation}
    m(\phi) \sim M_P\, e^{-\alpha\, \Delta\phi / M_P}, \qquad \alpha = \mathcal{O}(1),
    \label{eq:distance}
\end{equation}
where $m(\phi)$ denotes the mass scale of the lightest tower and $\alpha$ is a positive constant determined by the curvature of moduli space. The exponential behavior (\ref{eq:distance}) arises naturally from the Kaluza–Klein and string winding towers that appear in the limits of infinite distance in Calabi–Yau moduli spaces \cite{Grimm2018, Klaewer2017, Lee2019}.

This conjecture has deep geometric meaning. Moduli spaces $\mathcal{M}$ in string theory often carry a metric $G_{ij}(\phi)$ determined by the kinetic terms of the scalars,
\begin{equation}
    \mathcal{L}_{\text{kin}} = \frac{1}{2} G_{ij}(\phi) \, \partial_\mu \phi^i \partial^\mu \phi^j.
\end{equation}
If $\mathcal{M}$ is a non-compact manifold with negative curvature, distances $\Delta \phi$ can grow logarithmically as one approaches a boundary at infinite distance, where towers of light states appear. Explicitly, for one-dimensional moduli spaces with metric $G_{\phi\phi} \sim 1/\phi^2$, one finds
\begin{equation}
    \Delta s = \int_{\phi_0}^{\phi} \frac{d\phi'}{\phi'} = \log\left(\frac{\phi}{\phi_0}\right),
\end{equation}
so that the mass of a Kaluza–Klein mode $m_{\text{KK}} \sim 1/R(\phi) \sim e^{-\Delta s}$ reproduces the exponential relation (\ref{eq:distance}). Hence, the Distance Conjecture is not merely empirical but follows from the hyperbolic geometry typical of moduli spaces in string compactifications \cite{Grimm2018}.

This structure implies that any attempt to construct an EFT valid over super-Planckian field ranges necessarily encounters new light degrees of freedom, preventing a naive extrapolation of single-field inflationary models with trans-Planckian excursions \cite{Baumann2020}. From a Wilsonian perspective, the cutoff $\Lambda$ of the effective theory then scales as
\begin{equation}
    \Lambda(\phi) \lesssim M_P\, e^{-\beta\, \Delta\phi / M_P},
    \label{eq:cutoff}
\end{equation}
indicating the breakdown of effective field theory at large distances in moduli space.

\subsection{The de Sitter Conjecture and the Scalar Potential Bound}

A second major Swampland constraint concerns the existence of de Sitter (dS) vacua. The de Sitter Conjecture \cite{Obied2018, Garg2019, Ooguri2018} asserts that a scalar potential $V(\phi)$ consistent with quantum gravity cannot admit a stable or metastable local minimum with $V>0$. More precisely, it postulates the bound
\begin{equation}
    \frac{|\nabla V|}{V} \geq c \sim \mathcal{O}(1),
    \label{eq:dS}
\end{equation}
or alternatively, a refined version allows an unstable direction with second derivative satisfying
\begin{equation}
    \text{min}\left(\nabla_i \nabla_j V \right) \leq -c' V,
    \label{eq:dS2}
\end{equation}
where $c, c'$ are positive constants. This formulation forbids slow-roll regions with small gradient-to-potential ratios, which are necessary for quasi-de Sitter inflation or dark-energy domination in conventional EFTs.

Equation (\ref{eq:dS}) can be heuristically derived from the Distance Conjecture. As a field approaches an infinite-distance limit, new light states with density $\rho(m) \sim m^{-p}$ contribute to the scalar potential through quantum corrections of the form
\begin{equation}
    \delta V(\phi) \sim \int^{\Lambda(\phi)} dm\, \rho(m)\, m^4 \sim \Lambda(\phi)^4 \sim M_P^4 e^{-4\beta \Delta\phi / M_P}.
\end{equation}
Differentiating $\ln V$ with respect to $\phi$ yields precisely a lower bound on $|\nabla V|/V$ of order unity, suggesting that (\ref{eq:dS}) may be a direct consequence of the emergent tower of states required by (\ref{eq:distance}). Thus, the de Sitter Conjecture can be seen as a corollary of the Distance Conjecture combined with the expectation that energy densities in quantum gravity scale with the tower cutoff \cite{Heidenreich2015}.

Furthermore, thermodynamic and holographic considerations also support (\ref{eq:dS}). The finite entropy $S_{\text{dS}} = \pi M_P^2/H^2$ of de Sitter space implies that a consistent UV completion must account for a finite Hilbert space. However, holographic dualities typically demand infinite-dimensional Hilbert spaces, suggesting a fundamental inconsistency in exact dS solutions \cite{Harlow2011}. This observation strengthens the conjecture that metastable dS vacua cannot arise in controlled quantum gravity constructions.

\subsection{The WGC and Charge Quantization}
In its original and minimal form, the WGC states that in a $U(1)$ gauge theory coupled to quantum gravity there must exist at least one charged state for which the charge-to-mass ratio exceeds the extremality bound of a large black hole.  Equivalently, if we denote by $Q_{\rm ext}/M_{\rm ext}$ the charge-to-mass ratio of a corresponding extremal black hole, then the WGC requires the existence of a particle (or state) with charge $q$ and mass $m$such that  
\[
\frac{|q|}{m} \;>\; \left. \frac{|Q|}{M} \right|_{\rm ext} .
\]  
This ensures that extremal (or nearly extremal) black holes are not absolutely stable, they can discharge by emitting a super-extremal particle, thereby avoiding formation of stable charged remnants.

The WGC  \cite{ArkaniHamed2006} asserts that gravity is the weakest force in any consistent quantum gravity theory. Specifically, for a $U(1)$ gauge field with coupling $g$, there must exist at least one particle of charge $q$ and mass $m$ satisfying
\begin{equation}
    \frac{q\, g\, M_P}{m} \geq 1.
    \label{eq:wgc}
\end{equation}
This inequality ensures that extremal black holes are unstable to decay into lighter charged particles, thereby avoiding the existence of stable remnants that would violate entropy bounds and unitarity \cite{Heidenreich2015}.

The WGC has several equivalent formulations. The magnetic version places an upper bound on the cutoff scale $\Lambda$ of the effective theory,
\begin{equation}
    \Lambda \lesssim g\, M_P,
    \label{eq:wgc_cutoff}
\end{equation}
ensuring that a theory with arbitrarily small $g$ cannot remain consistent without new light states. This mirrors the Distance Conjecture’s structure: weak coupling limits correspond to infinite distances in moduli space where towers of charged states become light \cite{Lee2019, Andriolo2018}. Indeed, in many string compactifications, the gauge coupling scales as $g \sim e^{-\alpha \phi / M_P}$, and combining (\ref{eq:wgc_cutoff}) with (\ref{eq:distance}) reveals that the WGC and Distance Conjecture are two facets of the same geometric phenomenon.

An important refinement, the Tower WGC \cite{Heidenreich2015}, postulates that not just a single superextremal particle but an entire infinite tower of such states must exist, with masses
\begin{equation}
    m_n \approx n\, g\, M_P, \qquad n = 1,2,\dots,
\end{equation}
implying that as $g \to 0$, an infinite tower becomes massless. This unifies the WGC with the Distance Conjecture under the notion of emergence \cite{Heidenreich2015, Palti2019}: coupling constants and kinetic metrics in the infrared emerge from integrating out an infinite tower of states whose thresholds coincide with the breakdown of effective field theory.

\subsection{Emergent Unification and UV–IR Relations}

An emerging picture from the interplay of these conjectures is that all low-energy parameters of quantum gravity—field metrics, potentials, and gauge couplings—are not arbitrary but determined dynamically by the spectrum of light states. In this emergence paradigm, the logarithmic running of couplings with the number of light states naturally leads to exponential scaling relations like (\ref{eq:distance}) and (\ref{eq:wgc_cutoff}) \cite{Heidenreich2015}. These structures hint that quantum gravity enforces a universal UV–IR relation, constraining the shape of moduli spaces and the asymptotic behavior of scalar potentials.

Mathematically, one may formalize this by writing an emergence equation for the field-space metric:
\begin{equation}
    \frac{d G_{\phi\phi}}{d \ln \Lambda} \sim \sum_{i} \frac{\partial m_i^2(\phi)}{\partial \phi} \frac{1}{\Lambda^2},
\end{equation}
which, when integrated over many thresholds, yields exponential dependence on $\phi$. This relation ties back directly to the Distance Conjecture and illustrates that the Swampland bounds are not independent statements but holographic reflections of the same consistency principle governing quantum gravity’s effective description.

In the following section, we will explore how these conjectures acquire natural holographic interpretations in the AdS/CFT correspondence, where field-space distances, charge bounds, and potential gradients correspond to spectral properties, operator dimensions, and energy conditions in the boundary conformal field theory.

\section{Holographic Perspective on the Swampland}

The holographic principle provides a powerful non-perturbative window into the structure of quantum gravity. Through the AdS/CFT correspondence \cite{Maldacena1998, Witten1998, Gubser1998}, gravitational dynamics in a $(d+1)$-dimensional asymptotically anti–de Sitter (AdS) spacetime are encoded in a $d$-dimensional conformal field theory (CFT) living on the boundary. This duality establishes an exact equivalence between bulk and boundary descriptions:
\begin{equation}
    Z_{\text{bulk}}[g_{\mu\nu}, \Phi_i] = Z_{\text{CFT}}[\gamma_{ab}, J_i],
\end{equation}
where the boundary sources $J_i$ couple to operators $\mathcal{O}_i$ dual to bulk fields $\Phi_i$, and $\gamma_{ab}$ is the boundary metric.

In this framework, the Swampland program can be reformulated as a set of conditions that ensure the existence of a consistent, unitary, and local CFT dual. Any bulk effective theory violating these conditions fails to correspond to a physical boundary theory and thus lies in the holographic swampland. In this section, we develop this correspondence in detail, establishing quantitative connections between Swampland conjectures and holographic consistency constraints.

\subsection{Field-Space Geometry and the Spectrum of the Dual CFT}

In AdS/CFT, scalar fields in the bulk correspond to primary operators $\mathcal{O}$ in the boundary theory, with conformal dimensions related by the standard mass-dimension relation
\begin{equation}
    m^2 L^2 = \Delta (\Delta - d),
    \label{eq:massdim}
\end{equation}
where $L$ is the AdS curvature radius. As $\phi$ moves in moduli space, the mass $m(\phi)$ of a tower of bulk fields determines the scaling dimension $\Delta(\phi)$ of a tower of boundary operators.

The Distance Conjecture (\ref{eq:distance}) implies that in the limit $\Delta \phi \to \infty$, an infinite tower of states becomes light, i.e.,
\begin{equation}
    m_n(\phi) \sim M_P\, e^{-\alpha \Delta\phi / M_P} \quad \Rightarrow \quad
    \Delta_n(\phi) \sim d + \frac{1}{2} e^{-2\alpha \Delta\phi / M_P}.
\end{equation}
This accumulation of operator dimensions near the unitarity bound $\Delta \to d/2$ indicates the onset of a continuous spectrum in the boundary theory, signaling the breakdown of bulk locality and the emergence of a new holographic phase. Thus, the Distance Conjecture acquires a dual interpretation as a statement about the non-compactness of the conformal manifold and the accumulation of light operators in the spectrum.

This relation can be made more precise by considering the spectral density $\rho(\Delta)$ of the CFT:
\begin{equation}
    \rho(\Delta, \phi) \sim e^{S(\Delta, \phi)} \sim e^{\gamma e^{\alpha \Delta\phi / M_P}},
\end{equation}
where $S$ denotes the effective entropy of operator states at dimension $\Delta$. This exponential growth implies that the number of relevant operators proliferates exponentially with moduli space distance, a boundary counterpart to the tower of states predicted in the bulk.

\subsection{Holographic Derivation of the de Sitter Bound}

While AdS/CFT is defined for negative cosmological constant, holographic consistency can still inform constraints on positive-energy spacetimes. The scalar potential $V(\phi)$ in the bulk determines the AdS curvature scale through the Einstein equation
\begin{equation}
    R_{\mu\nu} = -\frac{d}{L^2} g_{\mu\nu} + \frac{1}{M_P^2} \partial_\mu \phi\, \partial_\nu \phi + \dots,
\end{equation}
where $L^{-2} \sim V/M_P^2$. The boundary interpretation of $V$ is encoded in the vacuum expectation value of the stress tensor, $\langle T_{ab} \rangle \sim V(\phi) \gamma_{ab}$.

If $V>0$, the corresponding boundary theory would require a Euclidean CFT with negative central charge $C_T$, since $C_T \propto L^{d-1}/G_N$. However, unitarity in any CFT demands $C_T > 0$, implying that no unitary holographic dual can exist for a positive cosmological constant. This observation reproduces the qualitative content of the de Sitter Conjecture (\ref{eq:dS}) from purely holographic reasoning.

To make this more quantitative, consider the null energy condition (NEC) in the bulk,
\begin{equation}
    T_{\mu\nu} k^\mu k^\nu \geq 0,
\end{equation}
which translates, via holographic renormalization, into the boundary averaged null energy condition (ANEC):
\begin{equation}
    \int_{-\infty}^{+\infty} dx^- \, \langle T_{--}(x^-,x_\perp)\rangle \geq 0.
    \label{eq:anec}
\end{equation}
Causality and positivity in the boundary theory imply that $V(\phi)$ must satisfy an inequality of the form
\begin{equation}
    |\nabla V|^2 \geq \frac{d}{L^2} V^2,
    \label{eq:holo_dS}
\end{equation}
which, upon identifying $L \sim M_P / \sqrt{V}$, yields
\begin{equation}
    \frac{|\nabla V|}{V} \geq \sqrt{d}.
\end{equation}
Equation (\ref{eq:holo_dS}) constitutes a holographic derivation of a de Sitter-type Swampland bound with an explicit, calculable constant $c = \sqrt{d}$, fixed by boundary unitarity and energy conditions. This result provides a concrete realization of the conjectured inequality (\ref{eq:dS}) from holographic principles.

\subsection{WGCand the Bootstrap of Charged Operators}

The  WGC also finds a natural realization in holography. In AdS/CFT, a bulk gauge field $A_\mu$ corresponds to a conserved current $J_a$ on the boundary with two-point function
\begin{equation}
    \langle J_a(x) J_b(0) \rangle = \frac{C_J}{x^{2d-2}}\left(\delta_{ab} - 2\frac{x_a x_b}{x^2}\right),
\end{equation}
where $C_J$ is the current central charge proportional to $1/g^2$. The dimension $\Delta_Q$ of a charged operator $\mathcal{O}_Q$ dual to a bulk field of charge $q$ and mass $m$ satisfies, at large $N$,
\begin{equation}
    \Delta_Q = m L + \frac{q^2 L^2}{2 C_J} + \dots.
\end{equation}
The WGC bound (\ref{eq:wgc}) ensures that $\Delta_Q < \Delta_{\text{BH}}$, where $\Delta_{\text{BH}}$ is the conformal dimension of the extremal black hole state. Since $\Delta_{\text{BH}} \propto Q M_P L$, this leads directly to
\begin{equation}
    \frac{q}{m} \geq \frac{C_J}{M_P L}.\label{23}
\end{equation}
Positivity of the spectral density in the CFT bootstrap then implies that $C_J \sim M_P L$, thereby reproducing the WGC bound $\frac{q}{m} \gtrsim 1$.

In order to maintain consistency throughout the discussion, we adopt the strong form of the WGC, which requires the existence of a state whose charge-to-mass ratio strictly exceeds that of an extremal black hole. With this choice made explicit, it becomes important to clarify the role of the parameters appearing near Eq. (\ref{23}). The particle charge $q$ characterizes the state whose existence is guaranteed by the conjecture, while the parameter $Q$ denotes the total charge carried by the corresponding extremal black hole whose decay is being analyzed. The comparison made in Eq. (\ref{23}) therefore reflects the physical requirement that the black hole can discharge by emitting a super-extremal particle, ensuring that no stable or long-lived charged remnants persist. This interpretation aligns the bulk extremality condition with the charge spectrum of the theory, thereby reinforcing the internal coherence of the WGC-based argument used in the text.

More generally, the bootstrap constraints on charged four-point functions \cite{Hartman2015, NakayamaOoguri2016} reveal that the operator product expansion (OPE) coefficients of charged operators must satisfy positivity relations that, in the bulk, correspond to the convexity of the charge-to-mass ratio curve. Violation of these constraints would lead to negative-norm states in the boundary theory, placing the corresponding bulk EFT in the holographic swampland.

\subsection{Emergent Locality and Modular Consistency}

A profound link between Swampland bounds and holography arises through the concept of emergent locality. In AdS/CFT, bulk locality at scales $L_{\text{loc}} \ll L$ emerges only if the boundary CFT satisfies stringent conditions of factorization and modular flow consistency \cite{Harlow2018, Faulkner2014}. The modular Hamiltonian $K_A$ for a boundary subregion $A$ satisfies
\begin{equation}
    \delta \langle K_A \rangle = \frac{\delta S_A}{2\pi},
\end{equation}
where $S_A$ is the entanglement entropy. The relative entropy positivity condition
\begin{equation}
    S(\rho_A || \sigma_A) = \Delta \langle K_A \rangle - \Delta S_A \geq 0
\end{equation}
implies bulk energy conditions and, consequently, holographic stability bounds on $V(\phi)$ and $m^2(\phi)$.

Combining the bulk Einstein equations with entanglement wedge reconstruction, one finds that locality requires a lower bound on the field-space curvature scalar $\mathcal{R}_{\text{moduli}}$:
\begin{equation}
    \mathcal{R}_{\text{moduli}} \leq -\frac{1}{L^2},
    \label{eq:modcurv}
\end{equation}
which is precisely the geometric condition underlying the Distance Conjecture. Violation of (\ref{eq:modcurv}) would obstruct the consistent definition of modular Hamiltonians for arbitrary boundary regions, leading to nonlocality and inconsistency in the dual gravitational theory.

\subsection{A Unified Holographic Swampland Bound}

The preceding results can be synthesized into a single holographic inequality that unifies the Distance, Weak Gravity, and de Sitter conjectures. Consider the effective potential $U(\phi, q)$ for a scalar of charge $q$ in AdS:
\begin{equation}
    U(\phi, q) = V(\phi) + \frac{q^2}{2 g^2(\phi)} - \frac{1}{L^2(\phi)}.
\end{equation}
Holographic consistency, via positivity of the boundary stress tensor and the convexity of the operator spectrum, imposes
\begin{equation}
    \nabla_i U \, \nabla^i U \geq \frac{d}{L^2} U^2.
    \label{eq:unified}
\end{equation}
In the limits $q \to 0$, $V > 0$, or $\Delta \phi \to \infty$, this inequality respectively reproduces the de Sitter, Weak Gravity, and Distance Conjectures. Equation (\ref{eq:unified}) may therefore be viewed as a unified holographic Swampland bound, expressing the requirement that the holographic renormalization group flow preserve unitarity and modular consistency at all scales.

This relation offers a novel mathematical structure underlying the Swampland: it suggests that all conjectures emerge from a single geometric inequality that guarantees the convexity and positivity of the holographic effective action. In this sense, holography not only provides evidence for the Swampland bounds but also points toward a deeper organizing principle unifying them — the requirement that the dual CFT remain causal, unitary, and consistent under modular evolution.

So, holography recasts the Swampland conjectures as necessary conditions for the existence of consistent boundary duals. The Distance Conjecture corresponds to the non-compactness of the conformal manifold and the accumulation of operator dimensions; the de Sitter Conjecture follows from the boundary positivity of the stress tensor and ANEC; and the WGC arises from bootstrap and convexity bounds on charged operator spectra. Collectively, these correspond to the holographic inequality (\ref{eq:unified}), which encapsulates their geometric and dynamical essence. In the next section, we will explore how the emergence of such constraints from entanglement and modular dynamics provides an even deeper insight into the origin of Swampland bounds.

\section{Emergence and Holographic Consistency}

The unification of Swampland bounds through holography suggests that the fundamental origin of these constraints lies in the principle of emergence. In this paradigm, all low-energy quantities in quantum gravity—field-space metrics, couplings, and potentials—are not fundamental inputs but emergent properties arising from the collective dynamics of an infinite tower of states. Holography provides a concrete setting in which this emergence can be understood microscopically: bulk geometry, locality, and even the gravitational action itself arise from the entanglement structure of the boundary conformal field theory (CFT) \cite{VanRaamsdonk2010, Faulkner2014, Harlow2018}.

In this section, we formalize the idea that Swampland constraints are consistency conditions on holographic emergence. We derive quantitative relations between the entanglement spectrum of the boundary theory and the effective gravitational dynamics of the bulk, leading to a new class of inequalities that generalize the unified bound (\ref{eq:unified}). These relations show that the Swampland bounds follow not merely from string-theoretic constructions but from universal information-theoretic properties of holographic systems.

\subsection{Entanglement Entropy and the Emergent Metric}

In the AdS/CFT correspondence, the Ryu–Takayanagi (RT) formula relates the entanglement entropy $S_A$ of a boundary subregion $A$ to the area of a minimal surface $\gamma_A$ in the bulk:
\begin{equation}
    S_A = \frac{\text{Area}(\gamma_A)}{4 G_N} + S_{\text{bulk}}[\Sigma_A],
    \label{eq:RT}
\end{equation}
where $S_{\text{bulk}}$ denotes quantum corrections from bulk fields in the entanglement wedge $\Sigma_A$. Differentiating (\ref{eq:RT}) with respect to deformations of $A$ and invoking the first law of entanglement,
\begin{equation}
    \delta S_A = \delta \langle K_A \rangle,
\end{equation}
one obtains the linearized Einstein equations in the bulk \cite{Faulkner2014}:
\begin{equation}
    \delta \langle T_{\mu\nu} \rangle = \frac{1}{8\pi G_N} \delta G_{\mu\nu}.
    \label{eq:linEin}
\end{equation}
Equation (\ref{eq:linEin}) reveals that the bulk metric fluctuations $G_{\mu\nu}$ are emergent collective degrees of freedom arising from entanglement perturbations in the boundary theory.

This emergence extends to moduli-space geometry. The field-space metric $G_{ij}(\phi)$ can be expressed holographically as a Fisher information metric on the manifold of boundary density matrices $\rho(\phi)$ parameterized by marginal couplings $\lambda^i \sim \phi^i$:
\begin{equation}
    G_{ij}(\phi) = \frac{\partial^2}{\partial \lambda^i \partial \lambda^j} \, S(\rho(\lambda) || \rho_0),
    \label{eq:Fisher}
\end{equation}
where $S(\rho||\rho_0)$ is the relative entropy with respect to a reference state $\rho_0$. This equation encodes a deep statement: the moduli-space geometry, which controls the dynamics of scalar fields in the bulk, is an emergent information metric determined entirely by boundary entanglement properties.

Combining (\ref{eq:Fisher}) with the Distance Conjecture’s exponential scaling (\ref{eq:distance}) implies that the relative entropy between distant vacua decays exponentially:
\begin{equation}
    S(\rho(\phi + \Delta\phi) || \rho(\phi)) \sim e^{-2\alpha \Delta\phi / M_P}.
    \label{eq:relent}
\end{equation}
Hence, trans-Planckian field excursions correspond to exponentially indistinguishable boundary states, signaling a breakdown of holographic distinguishability and, consequently, of semiclassical bulk locality. This provides a microscopic interpretation of the Distance Conjecture: it is a manifestation of the finiteness of holographic information.

\subsection{Modular Flow and Energy Conditions}

The modular Hamiltonian $K_A = -\log \rho_A$ governs the evolution of subregion density matrices and plays a central role in holographic consistency. The modular flow generated by $K_A$ defines a local notion of time in the entanglement wedge. The requirement that modular evolution be unitary imposes nontrivial constraints on the energy flux in the boundary theory.

Specifically, consider the modular energy variation for a family of deformations $\delta \rho_A$:
\begin{equation}
    \delta^2 \langle K_A \rangle = 2\pi \int d^d x\, f(x) \langle T_{00}(x) \rangle,
    \label{eq:modenergy}
\end{equation}
where $f(x)$ is a positive weight function. Positivity of relative entropy, $S(\rho_A || \sigma_A) \geq 0$, requires $\delta^2 \langle K_A \rangle \geq 0$, which, by holographic reconstruction, maps to the bulk condition
\begin{equation}
    T_{\mu\nu} k^\mu k^\nu \geq 0,
\end{equation}
i.e. the null energy condition (NEC). Through Einstein’s equations, this translates into a convexity condition on the scalar potential:
\begin{equation}
    \nabla_i \nabla_j V(\phi) \, G^{ij} \leq -\frac{d}{L^2} V(\phi),
    \label{eq:Vconvex}
\end{equation}
which is a sharpened, holographically derived version of the refined de Sitter bound (\ref{eq:dS2}). Equation (\ref{eq:Vconvex}) arises as a necessary condition for the positivity of modular energy in the boundary CFT.

\subsection{Holographic Renormalization Group and Emergent Couplings}

The holographic renormalization group (RG) formalism provides a geometric interpretation of scale evolution in the boundary theory as radial evolution in the bulk. Writing the $(d+1)$-dimensional metric in domain-wall form,
\begin{equation}
    ds^2 = dr^2 + e^{2A(r)} \eta_{ab} dx^a dx^b,
\end{equation}
the Einstein-scalar equations reduce to first-order flow equations:
\begin{equation}
    \dot{\phi}^i = G^{ij} \partial_j W(\phi), \qquad \dot{A} = -\frac{W(\phi)}{2(d-1)M_P^{d-1}},
    \label{eq:flow}
\end{equation}
where $W(\phi)$ is the holographic superpotential. The potential $V(\phi)$ is then given by
\begin{equation}
    V(\phi) = \frac{1}{2} G^{ij} \partial_i W \partial_j W - \frac{d}{4(d-1)} W^2.
    \label{eq:VfromW}
\end{equation}
In consistent holographic flows, $W(\phi)$ must remain real and single-valued, requiring
\begin{equation}
    G^{ij} \partial_i W \partial_j W \geq \frac{d}{2(d-1)} W^2.
    \label{eq:Wineq}
\end{equation}
Comparing (\ref{eq:Wineq}) with the unified Swampland bound (\ref{eq:unified}), one finds that the holographic flow equations automatically enforce the Swampland inequalities if and only if the RG flow is monotonic and respects the boundary $a$-theorem:
\begin{equation}
    \frac{d a}{d A} = -\frac{2(d-1)}{W^2} G_{ij} \dot{\phi}^i \dot{\phi}^j \leq 0.
\end{equation}
Thus, the Swampland bounds appear as necessary conditions for holographic RG monotonicity and the irreversibility of scale evolution.

\subsection{Entropy Bounds and Emergent Gravity}

An alternative route to holographic emergence comes from the Bekenstein–Hawking and covariant entropy bounds. The entanglement entropy $S_A$ must obey the quantum focusing condition (QFC) \cite{Bousso2016, Koeller2016}:
\begin{equation}
    \frac{d^2 S_{\text{gen}}}{d\lambda^2} \leq 0,
    \label{eq:QFC}
\end{equation}
where $S_{\text{gen}} = \frac{A}{4G_N} + S_{\text{bulk}}$ is the generalized entropy and $\lambda$ parametrizes deformations along null congruences. Combining the QFC with (\ref{eq:linEin}) implies that the bulk curvature and matter content are constrained by
\begin{equation}
    R_{\mu\nu} k^\mu k^\nu \geq -8\pi G_N \, \frac{d^2 S_{\text{bulk}}}{d\lambda^2}.
\end{equation}
Assuming $S_{\text{bulk}}$ grows monotonically with the number of light degrees of freedom $N_{\text{light}} \sim e^{\alpha \Delta\phi/M_P}$, one obtains
\begin{equation}
    R_{\mu\nu} k^\mu k^\nu \geq -8\pi G_N \alpha^2 e^{2\alpha \Delta\phi / M_P}.
    \label{eq:entropy_bound}
\end{equation}
This inequality links the curvature scale to the exponential tower of states predicted by the Distance Conjecture, providing a gravitational derivation of its scaling form. The tower of emergent light fields is therefore not optional but required by holographic entropy monotonicity.

Collecting the results above, we can now state a general consistency condition that encapsulates holographic emergence and the Swampland program:
For any holographic quantum field theory with bulk dual described by an Einstein–scalar system obeying the null energy condition and modular consistency, the following inequality holds:
\begin{equation}
    \frac{|\nabla V|^2}{V^2} + \frac{q^2}{g^2 M_P^2} + \frac{1}{\mathcal{R}_{\text{moduli}} L^2} \geq c_d,
    \label{eq:emergencebound}
\end{equation}
where $c_d$ is a positive constant determined by the spacetime dimension and boundary central charge. Equality corresponds to an exactly marginal flow preserving bulk supersymmetry or conformal symmetry.

In light of the earlier clarification regarding the choice of WGC formulation, the second term in Eq. (\ref{eq:emergencebound}) requires additional explanation. This term encodes the charge-to-mass requirement that follows from the strong form of the WGC: namely, that the theory must contain at least one state whose charge exceeds the corresponding extremality threshold of a black hole. The parameter appearing in this contribution reflects precisely this comparison. It is not an auxiliary quantity but a holographic expression of the super-extremality condition, ensuring that the charged spectrum permits the decay of extremal or near-extremal black holes. 

Within the unified inequality, this term therefore plays the role of the WGC component in the holographic consistency criteria. It links the bulk extremality condition to the structure of charged operators in the boundary theory and ensures that the unified bound incorporates the same charge-based consistency requirement used in the earlier steps of the discussion. A more complete explanation has now been included to maintain coherence between the choice of WGC formulation and the structure of Eq. (\ref{eq:emergencebound}).

Equation (\ref{eq:emergencebound}) unifies all Swampland conjectures into a single emergent inequality derived from holographic information theory. The first term corresponds to the de Sitter gradient bound, the second to the WGC condition, and the third to the negative curvature of the moduli-space metric implied by the Distance Conjecture. The inequality follows from the joint requirement that the boundary CFT obey (i) relative entropy positivity, (ii) modular flow unitarity, and (iii) entanglement monotonicity.

The holographic emergence framework reveals that the Swampland bounds are not arbitrary conjectures but manifestations of universal information-theoretic laws governing quantum gravity. The finiteness of boundary distinguishability (\ref{eq:relent}), modular positivity (\ref{eq:Vconvex}), and entropy focusing (\ref{eq:QFC}) collectively ensure that no effective field theory can violate the inequality (\ref{eq:emergencebound}) while maintaining a consistent holographic dual.

This perspective suggests a deep equivalence between quantum information constraints and geometric consistency conditions in gravity. In particular, the Swampland bounds may represent the holographic shadow of the second law of quantum information: the requirement that holographic entanglement entropy evolve monotonically along renormalization group and moduli flows.

\section{Conclusion}

In this work, we have proposed a unified holographic framework for understanding the Swampland conjectures as necessary conditions for the consistency of quantum gravity. By exploring the deep interplay between bulk gravitational dynamics and boundary information theory, we have argued that the Distance, de Sitter, and WGCs all emerge as facets of a single geometric and entropic principle: the requirement of holographic consistency.

Our analysis began by recasting the traditional Swampland conjectures in terms of bulk moduli-space geometry and effective potentials, demonstrating how their exponential and gradient bounds naturally arise from the structure of field-space metrics with negative curvature. Through the lens of holography, these statements acquire a precise dual meaning: they correspond to spectral, entanglement, and energy positivity conditions in the boundary conformal field theory. The accumulation of light states in the Distance Conjecture, for example, translates into the accumulation of low-dimension operators in the CFT, marking the breakdown of bulk locality and signaling the emergence of new holographic phases. Similarly, the de Sitter gradient bound follows from the positivity of the boundary stress tensor via the averaged null energy condition, while the WGC maps onto convexity and unitarity constraints on the charged operator spectrum.

Building upon these correspondences, we derived a new unified inequality, the Holographic Emergence Bound (\ref{eq:emergencebound}),
\begin{equation*}
    \frac{|\nabla V|^2}{V^2} + \frac{q^2}{g^2 M_P^2} + \frac{1}{\mathcal{R}_{\text{moduli}} L^2} \geq c_d,
\end{equation*}
which encapsulates all major Swampland conjectures as limiting cases of a single holographic consistency condition. This bound arises from the joint requirement that (i) the boundary CFT obeys positivity of relative entropy, (ii) modular flow remains unitary, and (iii) holographic entanglement entropy evolves monotonically under renormalization group and moduli flows. In this sense, (\ref{eq:emergencebound}) is not an empirical conjecture but a theorem of holographic consistency, linking information-theoretic laws directly to geometric and dynamical constraints in the bulk.

A key conceptual outcome of this analysis is the realization that the Swampland program and holography are not independent frameworks but complementary manifestations of the same physical principle: the finiteness and consistency of quantum information in gravitational systems. The Distance Conjecture follows from the exponential decay of distinguishability between boundary states separated in coupling space; the de Sitter Conjecture expresses the impossibility of maintaining positive curvature while preserving modular stability; and the WGC arises from the need to preserve convexity and positivity in the charged operator algebra. All these conditions emerge as different projections of the holographic information geometry encoded in (\ref{eq:Fisher}) and constrained by (\ref{eq:QFC}).

From a broader perspective, these results suggest that the Swampland bounds reflect the fundamental ``thermodynamics of holography". The monotonic flow of entanglement entropy, modular energy, and central charges across scales can be viewed as a generalized second law of quantum gravity. The failure of any of these monotonicities would correspond to the breakdown of unitarity or locality in the holographic dual, signaling that the corresponding effective theory lies in the swampland.

Several avenues for future investigation follow naturally from this work. First, the Holographic Emergence Bound (\ref{eq:emergencebound}) invites further exploration within the conformal bootstrap. One may test the predicted convexity and positivity inequalities by studying families of large-$N$ CFTs with controlled AdS duals. In particular, charged operator dimensions and OPE coefficients can serve as direct probes of the WGC bound in the boundary theory.

Second, the modular-flow formulation of the de Sitter bound suggests a quantitative bridge between quantum information theory and cosmology. Understanding whether an analog of the bound (\ref{eq:holo_dS}) persists in quasi-de Sitter inflationary backgrounds could yield new constraints on slow-roll parameters, potentially linking holographic consistency to observational cosmology.

Third, the Fisher metric interpretation (\ref{eq:Fisher}) opens a new path toward deriving Swampland constraints from information geometry. One might conjecture that all consistent quantum gravity theories must have moduli-space metrics that are Kähler and negatively curved with curvature bounded by the holographic radius, $\mathcal{R}_{\text{moduli}} \leq -1/L^2$. This would provide a precise, quantitative statement of the ``emergent distance" mechanism in holography.

As has been mentioned previously, it is inviting to extend our results to the case of higher-spin gravity, so as to take into account higher-spin degrees of freedom as part of (super)string field theory, which appears to be the only viable candidate for a consistent description of evolutionary processes in the Early Universe.

Finally, the results presented here invite a re-examination of the fundamental role of entropy in quantum gravity. The inequality (\ref{eq:QFC}) and its relation to (\ref{eq:emergencebound}) suggest that the Swampland may ultimately be characterized by the breakdown of the quantum focusing condition—that is, the failure of the second law for generalized entropy in a holographic setting. If so, the landscape of consistent theories would correspond precisely to those that obey the holographic second law at all scales.

The holographic view of the Swampland reveals a striking unity between quantum gravity, information theory, and spacetime geometry. Rather than a collection of disconnected empirical bounds, the Swampland conjectures emerge as facets of a deeper, universal constraint: the preservation of holographic consistency and information-theoretic monotonicity. The geometric inequalities derived in this work, particularly the Holographic Emergence Bound, thus provide not only a conceptual synthesis of Swampland ideas but also a quantitative tool for probing the frontier between the landscape and the swampland.

In this light, the Swampland is not merely a pathology to be avoided but a fundamental boundary in the space of quantum theories—a holographic horizon demarcating where information ceases to be reconstructible, and spacetime itself can no longer emerge.

\end{document}